\documentclass[12pt]{article}
\usepackage{a4wide}
\usepackage{epsfig}
\usepackage{amsmath}

\newlength{\absize}
\catcode`@=11
\def\citer{\@ifnextchar [{\@tempswatrue\@citexr}{\@tempswafalse\@citexr[]}}

\def\@citexr[#1]#2{\if@filesw\immediate
  \write\@auxout{\string\citation{#2}}\fi
  \def\@citea{}\@cite{\@for\@citeb:=#2\do
    {\@citea\def\@citea{--\penalty\@m}\@ifundefined
       {b@\@citeb}{{\bf ?}\@warning
       {Citation `\@citeb' on page \thepage \space undefined}}%
\hbox{\csname b@\@citeb\endcsname}}}{#1}}
\catcode`@=12

\begin{document}
  \thispagestyle{empty}
  \pagestyle{empty}
  \renewcommand{\thefootnote}{\fnsymbol{footnote}}
\newpage\normalsize
    \pagestyle{plain}
    \setlength{\baselineskip}{4ex}\par
    \setcounter{footnote}{0}
    \renewcommand{\thefootnote}{\arabic{footnote}}
\newcommand{\preprint}[1]{%
  \begin{flushright}
    \setlength{\baselineskip}{3ex} #1
  \end{flushright}}
\renewcommand{\title}[1]{%
  \begin{center}
    \LARGE #1
  \end{center}\par}
\renewcommand{\author}[1]{%
  \vspace{2ex}
  {\Large
   \begin{center}
     \setlength{\baselineskip}{3ex} #1 \par
   \end{center}}}
\renewcommand{\thanks}[1]{\footnote{#1}}
%\begin{flushright}
%March 2009
%\end{flushright}
%\begin{flushright}
%Revised version of arXiv: 0810.4219
%\end{flushright}
\vskip 0.5cm

\begin{center}
{\large \bf Testing Spatial Noncommutativity via Magnetic Hyperfine
Structure Induced by Fractional Angular Momentum of Rydberg System}
%Verifications of Quantum Effects of Spatial Noncommutativity}
%
%{\large \bf Induced Magnetic Hyperfine Structures by Fractional Zero
%Point Angular Momentum Emerging from Spatial Noncommutativity}
%and Aharonor-Bohm System}
% in Noncommutative Space}
\end{center}
\vspace{1cm}
%%%%%%
\begin{center}
Jian-Zu ZHANG$\;^{a, {\ast}}$, Hong-Ping LIU$\;^b$, Wei CAO$\;^c$,
Ke-Lin GAO$\;^b$
\end{center}
%-----------------------------------
%   Address
%-----------------------------------
%\vspace{1cm}
\begin{center}
$^a$ Institute for Theoretical Physics, East China University of
Science and Technology, Box 316, Shanghai 200237, People's Republic
of China\\
$^b$ State Key Laboratory of Magnetic Resonance and Atomic and
Molecular Physics, Wuhan Institute of Physics and Mathematics,
Chinese Academy of Science, Wuhan 430071, People's Republic of China\\
$^c$ Department of Physics, University of Fribourg; Ch du Musee 3,
1700 Fribourg, Switzerland
\end{center}
%\vspace{1cm}
%%%%%%%%%%%%%%%%%%%%%%%%%%%%%%%%%%%%%%%%%%%%%%%%%%%%%%%%%%%%%%%

\begin{flushleft}


\end{flushleft}

%%%%%%%%%%%%%%%%%%%%%%%%%%%%%%%%%%%%%%%%%%%%%%%%%%%%%%%%%%%%%%%

\begin{abstract}

An approach to solve the critical problem of testing quantum effects
of spatial noncommutativity is proposed.
Magnetic hyperfine structures in a Rydberg system induced by
fractional angular momentum originated from spatial noncommutativity
are discussed.
The orders of the corresponding magnetic hyperfine splitting of
spectrum $\sim 10^{-7} -  10^{-8} eV$ lie within the limits of
accuracy of current experimental measurements.
Experimental tests of physics beyond the standard model are the
focus of broad interest. We note that the present approach is
reasonable achievable with current technology.
The proof is based on very general arguments involving only the
deformed Heisenberg-Weyl algebra and the fundamental property of
angular momentum.
Its experimental verification would constitute an advance in
understanding of fundamental significance,
and would be a key step towards a decisive test of spatial
noncommutativity.

\end{abstract}

\begin{flushleft}
$^{\ast}$ Correspondence should be addressed to: J.-Z. Zhang
%Corresponding author. \\
%$\;$
(email: jzzhang@ecust.edu.cn)
%(J.-Z. Zhang).

%Correspondence should be addressed to: J.-Z. Zhang
%(jzzhang@ecust.edu.cn).
\end{flushleft}
\clearpage

%%%%%%%%%%%%%%%%%%%%%%%%%%%%%%%%%%%%%%%%%%%%%%%%%%%%%%%%%%%%%%%

{\bf 1. Introduction} --
%
%\vspace{0.5cm}
%
As one of the current candidates in tracking down new physics beyond
the standard model, quantum mechanics in noncommutative space
(NCQM)~\citer{Witt86,JZZ08} should be verifiable.~\footnote %bbb
%%%%%%%%%%%%%%%%%%%%%%%%%%%%%%%%%%%%%%%%%%%%%%%%%%%%%%%%%%%%%%%%%%%
{This paper focuses on the low energy relics of noncommutative
quantum theory and construct formalism, which closely relates to a
way testable by current experiments. It is enough to work in
deformed formalism at the NCQM level.}
%%%%%%%%%%%%%%%%%%%%%%%%%%%
Modifications of spatial noncommutativity (NC) to normal quantum
theory depending on vanishingly small NC parameters, which lead to
NC quantum effects are usually far beyond experimental accuracy.
Therefore, a widely held view is that NCQM can only make predictions
outside the range of experimental observation.
%If true, then it would become necessary to abandon it as testable
%science.
%
However, the conclusion is premature~\cite{CHLZ}.
Indeed, attempts in recent experiments performed by Connerade et
al.~\cite{CHLZ}
%and others~\cite{JZZ08}
suggest that there may be a way to test for NCQM.

Recently, it has been found~\cite{JZZ04b,JZZ06,JZZ08} that the
vanishingly small NC constants~\cite{CHKLO,BRACZ}, which usually
appear in NC corrections of any physical observable, cancel out in
the fractional angular momentum (FAM) originated from spatial
noncommutativity under well-defined conditions. It turns out that
FAM results in the unusual zero-point value $\hbar/4$.
This provides a distinct signature of spatial noncommutativity,
which survives into the normal quantum scale.
The difficulty involved in testing spatial noncommutativity via FAM
is that direct measurements of FAM are a challenge enterprise.

%%%%%%%%%%%%%%%%%%%%%%%%%%%%%%%%%%%%%%%%%%%%%%%%%%%%%%%%%%%%%%%%%%%%
With particular emphasis on feasible experimental tests, this paper
proposes an approach of testing spatial noncommutativity via
measuring magnetic hyperfine structures (MHFS~\cite{Slat,Yang})
induced by FAM in a Rydberg system.
The orders of the corresponding splitting of MHFS
$\sim 10^{-7} - 10^{-8} eV$
lie within the limits of accuracy of current experimental
measurements, and can be detected by using existing technology.
%%%%%%%%%%%%%%%%%%%%%%%%%%%%%%%%%%%%%%%%%%%%%%%%%%%%%%%%%%%%%%%%%%%
The significant advance of the proposed method is that it
%substantially
%extends
%advances quantum theory in NC space,
solves a critical outstanding problem of NC quantum effects being
unmeasurable, paves the way for notable progress and will lead to
the first real test of spatial noncommutativity. Our proof is based
on a very general argument involving only the deformed
Heisenberg-Weyl algebra and the fundamental property of angular
momentum. Therefore, if it is achieved experimentally, this will
constitute an advance in understanding of fundamental significance.

%\vspace{0.5cm}

%%%%%%%%%%%%%%%%%%%%%%%%%%%%%%%%%%%%%%%%%%%%%%%%%%%%%%%%%%%%%%%

{\bf 2. Review of FAM originated from spatial
noncommutativity}~\cite{JZZ04b,JZZ04a,JZZ06,JZZ08} --
%
%\vspace{0.5cm}
%
We investigate ion motion in the laboratory system, trapped in a
uniform magnetic field ${\bf B}$ aligned along the $z$-axis and an
electrostatic potential~\cite{JZZ08}
%The mass and charge of the ion are $m_I$ and $q^*=Z^* e(>0)$.
%%%
\begin{equation}
\label{Eq:V}%1e
V_{eff} = V_{eff,2}+V_{eff,z}
= \frac{m_I}{2}(\omega_{\rho}^2x_ix_i + \omega_z^2z^2),
\end{equation}
%%%
(the summation convention is used, $i, j=1,2$), where $m_I$ is ion
mass,
$\omega_{\rho}$ and
$\omega_z$
are characteristic frequencies, respectively, in the
$(x_1,x_2)$-plane and $z$ direction.
The vector potential $A_i$ of ${\bf B}$ is chosen as
%
%\begin{equation}
%\label{Eq:A}%?e
$A_i=-B\epsilon_{ij}x_j/2$, $A_z=0.$
%\end{equation}
%%%%%%
The Hamiltonian $H(x,p)$ of the trapped ion can be decomposed into
%%%
%\begin{equation}
%\label{Eq:H}%?e
$H=H_2+H_z$,
%\end{equation}
%%%
%\begin{equation}
%\label{Eq:Hz}%?e
where $H_z(z,p_z) = p_z^2/2m_I + m_I\omega_z^2 z^2/2$, and
%is a harmonic Hamiltonian, and
%
\begin{eqnarray}
\label{Eq:H2}%2e
H_2(x,p) = H_{k,2}+V_{eff,2}
%
%= \frac{1}{2m_I}(p_i-q^*A_i)(p_i-q^*A_i)
%%
%+ \frac{m_I}{2}\omega_{\rho}^2x_ix_i
%\nonumber\\
%
&=& \frac{1}{2m_I}p_i^2+\frac{1}{2}\omega_c\epsilon_{ij} p_i
x_j+\frac{1}{2}m_I\omega_P^2 x_i^2,
\end{eqnarray}
where $H_{k,2} = \sum_i(p_i-q^*A_i)^2/2m_I$ is the mechanical
kinetic energy operator which is different from the canonical
kinetic energy operator $p_ip_i/2m_I$. $\;$
$H_2$ is a two dimensional Chern-Simons Hamiltonian with the
cyclotron frequency $\omega_c=q^*B/m_I$, effective charge $q^*=Z^*
e(>0)$ and the characteristic frequency
$\omega_P=(\omega_{\rho}^2+\omega_c^2/4)^{1/2}$.
In the following, we focus on the $H_2$.

%%%%%%
The deformed Hamiltonian $H_2(\hat x,\hat p)$ in noncommutative
space can be obtained by reformulating the corresponding undeformed
$H_2(x,p)$ in terms of deformed canonical variables $\hat x_i$ and
$\hat p_i$
%
%The deformed canonical variables $\hat x_i$ and $\hat p_i$
which satisfy two dimensional deformed Heisenberg-Weyl algebra
%%%
%\begin{equation}
%\label{Eq:xp}%e
$$[\hat x_i,\hat x_j]=i\xi^2\epsilon_{ij}\theta,\;
[\hat x_i,\hat p_j]=i\hbar\delta_{ij},\;
[\hat p_i,\hat p_j]=i\xi^2\epsilon_{ij}\eta,$$
%\end{equation}
%%%
where $\theta$ and $\eta$ are the constant parameters of spatial
noncommutativity, independent of position and momentum;
$\epsilon_{ij}$ is a two-dimensional antisymmetric unit tensor with
$\epsilon_{12}=-\epsilon_{21}=1,$ $\epsilon_{11}=\epsilon_{22}=0$.
The scaling factor $\xi$ is defined as
%%%%%%
%\begin{equation}
%\label{Eq:xi-1}%?e
$\xi=(1+\theta\eta/4\hbar^2)^{-1/2}.$
%\end{equation}
%%%%%%

The deformed Heisenberg-Weyl algebra
%Eqs.~(\ref{Eq:xp})
can be realized by
%undeformed variables
$x_{i}$ and $p_{i}$ as follows:
%%%
%\begin{equation}
%\label{Eq:hat-x-p}%e
$$\hat x_{i}=\xi(x_{i} - \frac{1}{2\hbar}\theta\epsilon_{ij}p_{j}),\;
\hat p_{i}=\xi(p_{i} + \frac{1}{2\hbar}\eta\epsilon_{ij}x_{j}),$$
%\end{equation}
%%%
where $x_{i}$ and $p_{i}$ satisfy the undeformed Heisenberg-Weyl
algebra
%\begin{equation}
%\label{Eq:x-p}%e
$[x_{i},x_{j}]=[p_{i},p_{j}]=0,\; [x_{i},p_{j}]=i\hbar\delta_{ij}.$
%\end{equation}
%
%Using Eqs.~(\ref{Eq:hat-x-p}), t
The deformed $H_2(\hat x,\hat p)$ can be further expressed by $x_i$
and $p_i$ as $\hat H_2(x,p)$:
\begin{eqnarray}
\label{Eq:CS-H1}%3e
\hat H_2(x,p) &=& \hat H_{k,2}(x,p) + \hat V_{eff,2}(x)
\equiv \frac{1}{2M}(p_i+\frac{1}{2}G\epsilon_{ij} x_j)^2
+\frac{1}{2}K x_i^2
\nonumber\\
&=& \frac{1}{2M} p_i^2+\frac{1}{2M}G\epsilon_{ij} p_i
x_j+\frac{1}{2}M\Omega_P^2 x_i^2,
\end{eqnarray}
where the effective parameters %$M, \Omega_c, \Omega_p$
$M, G, \Omega_P$ and $K$ are defined as
%%%
\begin{eqnarray}
\label{Eq:M-K}%e
\frac{1}{2M}&\equiv&\xi^2(\frac{1}{2m_I}c_1^2
+ \frac{1}{16\hbar^2}m_I\omega_{\rho}^2\theta^{\;2}),\;
\frac{G}{2M}\equiv\xi^2(\frac{1}{m_I}c_1 c_2
+ \frac{1}{4\hbar}m_I\omega_{\rho}^2\theta),
\nonumber\\
M\Omega_P^2&\equiv&\xi^2(\frac{1}{m_I}c_2^2
+ \frac{1}{2}m_I\omega_{\rho}^2),\quad\quad\quad\quad
K\equiv M\Omega_P^2 - \frac{1}{4M}G^2,
\nonumber
\end{eqnarray}
%%%
%$1/2M\equiv\xi^2(c_1^2/2m_I+\kappa\theta^{\;2}/8\hbar^2)$,
%%
%$G/2M\equiv\xi^2(c_1 c_2/m_I + m_I\omega_{\rho}^2\theta/2\hbar)$,
%%\nonumber\\
%%
%$M\Omega_P^2\equiv\xi^2(c_2^2/m_I + m_I\omega_{\rho}^2)$,
%%
%$K\equiv M\Omega_P^2-G^2/4M$,
%
and
$c_1=1+m_I\omega_c\theta/4\hbar,\;c_2=m_I\omega_c/2+\eta/2\hbar.\;$
$\hat H_2$ can be changed into two uncoupled harmonic
modes~\cite{JZZ04b,JZZ08}.

Similarly, the deformed angular momentum
%$\hat J_z$
$J_z(\hat x,\hat p)=\epsilon_{ij}\hat x_i\hat p_j$ can be expressed
by undeformed variables $x_i$ and $p_i$ as
$$\hat J_z(x,p) = \epsilon_{ij}x_ip_j
- \frac{1}{2\hbar}\xi^{2}\left(\theta p_i p_i +\eta x_i x_i\right).$$
%%%
%\begin{equation}
%\label{Eq:J1}%e
%\hat J_z(x,p) = \epsilon_{ij}x_ip_j
%%
%- \frac{1}{2\hbar}\xi^{2}\left(\theta p_i p_i +\eta x_i x_i\right).
%\end{equation}
%%%
The corrections due to spatial noncommutativity are terms
$O(\theta)$ and/or $O(\eta)$, which lead to $\hat J_z$ taking
fractional value.
The existing upper bounds of $\theta$ and $\eta$ are
$\theta/(\hbar c)^2\le (10 \;TeV)^{-2}$ \cite{CHKLO}
and
$|\sqrt{\eta}\,|\le 1 \mu eV/c$ \cite {BRACZ}.
$O(\theta)$ and $O(\eta)$ are vanishingly small, so that the
corrections of spatial noncommutativity are beyond the limits of
measurable accuracy of experiments.
%

%%%%%%%%%%%%%%%%%%%%%%%%%%%%%%%%%%%%%%%%%%%%%%%%%%%%%%%%%%%%%%%%%%%

{\bf 3. Reduction for massive system} --
We found a testable effect of spatial noncommutativity in the
reduced system of $\hat H_2$.
Because $\hat H_2$ and $\hat H_{k,2}$ do not commute,
different from the massless model considered in~\cite{DJT}, the
difficulty of reduction for the massive model is how to treat the
mechanical kinetic energy $\hat H_{k,2}$. To get rid of this
difficulty, the reducing procedure is adopted in the following
steps.

The ion oscillates harmonically with an axial frequency
%$\omega_z$
along the z-axis (its energy alternates between kinetic and
potential energy). In the (1, 2)-plane, it executes a superposition
of a fast circular cyclotron motion of an effective cyclotron
frequency
%$G/M$
%a cyclotron frequency $\omega_c$
with a small radius (its energy is almost exclusively kinetic
energy), and a slow circular magnetron drift motion of an effective
magnetron frequency
%$\Omega_m\equiv \omega_z^2/(2G/M)$
%a magnetron frequency $\omega_m\equiv \omega_z^2/2\omega_c$
in a large orbit (its energy is almost exclusively potential
energy).
%
%The  potential
$\hat V_{eff,2}$ is reduced by reducing the amplitudes of the
radio-voltage and the $dc$ voltage applied between the electrodes of
the ring and two end caps of the combined trap.
%
%We use an appropriate laser trapping and cooling field to slow the
%energy of the ion down to the ground state
%%lowest eigenvalue
%of $\hat H_2$.
%%~\cite{SST}.
%
We use, e.g., Doppler cooling to slow the energy of ion down to the
$mK$ and then cool the ion to the ground state of $\hat H_2$ with
the sideband cooling~\cite{Sten,WI97}.
By synchronizing the laser field with $\hat V_{eff,2}$ reduction,
the ion is kept in the ground state of the reducing $\hat H_2$.
In $\hat V_{eff,2}\to 0$ the axial and the magnetron-like motions
disappear, only the cyclotron motion survives.
%
%Thus $\hat H_2 \to \hat {\mathcal{H}}_2 = \hat H_{k,2}$,
Thus $\hat H_2 \to \hat H_2^{(\hat {V}\to 0)} = \hat H_{k,2}$, and
the energy of the survived
%cyclotron
motion is the ground value $\hat {\mathcal{E}}_{k,0}$.
%of $\hat H_{k,2}$.
%
%The ion is stabilized in this ground state
%%of $\hat H_{k,2}=\hat {\mathcal{E}}_{k,0}$
%by laser fields.
%%trapping and cooling
%%field during the entire subsequent proccess.
%

%%%%%%%%%%%%%%%%%%%%%%%%%%%%%%%%%%%%%%%%%%%%%%%%%%%%%%%%
%Keeping $\hat H_{k,2}=\hat {\mathcal{E}}_{k,0}$,
Taking $\hat H_2^{(\hat {V}\to 0)} = \hat {\mathcal{E}}_{k,0}$ as
the initial condition, the reduced system is obtained by resetting
an electric field
$\tilde {\bf E}$ of harmonic potential $m_I(\tilde
{\omega}_{\rho}^2x_ix_i + \tilde {\omega}_z^2z^2)/2$,
which leads to a full Hamiltonian
$\tilde H_2=\hat H_{k,2} + \tilde K x_ix_i/2$
(in Eq.~(3)
%Eq.~(6) of \cite{JZZ08}
we replace $\omega_{\rho}$ with $\tilde {\omega}_{\rho}$, then $K$
and $G$
%$G, \Omega_P$ and $K$
are replaced with $\tilde K$ and $\tilde G$).
%$\tilde G, \tilde \Omega_P$ and $\tilde K$).
%The electric field
$\tilde {\bf E}$ satisfies the condition that the ion is trapped in
the first stability range of the Paul trap. Thus $\tilde {\bf E}$ is
weak.
%
%The magnetic field
%$\bf B$ is strong enough, thus
The original $\bf B$ is fixed such that the corresponding energy
interval $\Delta \hat {\mathcal{E}}_{k} = \hat {\mathcal{E}}_{k,1} -
\hat {\mathcal{E}}_{k,0}$ is large enough so that $\tilde {\bf E}$
cannot disturb the ion from the ground state $|\hat
{\mathcal{E}}_{k,0}\rangle$ to the first excited state $|\hat
{\mathcal{E}}_{k,1}\rangle$ of $\hat H_2^{(\hat {V}\to 0)}$.
Thus the system remains in the ground state.
In the subspace $\{|\hat {\mathcal{E}}_{k,0}\rangle_i\}$ of the
ground state, for any state $|\psi\rangle = \sum_i c_i |\hat
{\mathcal{E}}_{k,0}\rangle_i$ we obtain
%
%$\hat H_{k,2}|\psi\rangle = \hat {\mathcal{E}}_{k,0}|\psi\rangle$.
%
$\tilde H_{2}|\psi\rangle
= (\hat H_{k,2} + \tilde K x_ix_i/2)|\psi\rangle
%
%= \hat H_{k,2}|\psi\rangle + (\tilde K x_ix_i/2)|\psi\rangle
%
= (\hat {\mathcal{E}}_{k,0} + \tilde K x_ix_i/2)|\psi\rangle$.
%
%the quantum behavior of $\hat H_2^{(V=0)} = \hat H_{k,2}$ is like a
%constant operator
%
Therefore, in the subspace $\{|\hat {\mathcal{E}}_{k,0}\rangle_i\}$
of the ground state, $\tilde H_{2}$ is reduced to:
%The reduced Hamiltonian is read out from this equation:
%
\begin{equation}
\label{Eq:H20}%4e
\tilde H_{2} \to
\hat {\mathcal{E}}_{k,0} + \frac{1}{2}\tilde K x_ix_i
\equiv  \tilde H_{2}^{(0)}.
\end{equation}
%(definition of $\tilde K$, see Eq.(6) of \cite{JZZ08}).

%%%%%%%%%%%%%%%%%%%%%%%%%%%%%%%%%%%%%%%%%%%%%%%%%%%%%%%%%%%%%%%%%%%%%%%
%\textit{Constraints} -- From $\tilde L_2^{(0)}$
%%$\hat L_2^{(0)}$
%(Eq.(17) of \cite{JZZ08}, the general expression of the reduced
%Lagrangian in a magnetic field) we obtain
%%
%$\tilde {H_{2}^\prime}^{(0)} \equiv p_i\dot{x_i}
%%
%- \tilde L_2^{(0)} = \tilde H_{2}^{(0)}$.
%%$\hat {H_{2}^\prime}^{(0)} \equiv p_i\dot{x_i}
%%%({\partial \hat L_2^{(0)}}/{\partial x_i})p_i
%%- \hat L_2^{(0)}$
%%
%It proves that $\tilde H_{2}^{(0)}$ is correct.
%%({\partial \hat L_2^{(0)}}/{\partial x_i})p_i
%- \hat L_2^{(0)}$
%$\tilde H_{2}^{(0)}$.
%
%This system is a constrained one.
%It indicates that
%%
%$\hat {H_{2}^\prime}^{(0)} = \hat H_{2}^{(0)}$.
%
%This proves correctness of $\hat H_{2}^{(0)}$.
%%%%%%%%%%%%%%%%%%%%%%%%%%%%%%%%%%%%%%%%
The reduced system $\tilde H_{2}^{(0)}$ is a constrained
one~\cite{JZZ08}.
%
%%%%%%
The Lagrangian corresponding to $\tilde H_{2}$ is
$\tilde L_2 = M\dot{x_i}\dot{x_i}/2
+ \tilde G\epsilon_{ij}\dot{x_i}x_j/2 -\tilde K x_i x_i/2$.
The reduced Lagrangian corresponding to $\tilde H_{2}^{(0)}$ is
$\tilde L_2^{(0)}=\tilde G\epsilon_{ij}\dot{x_i}x_j/2 -
\tilde K x_i x_i/2 - \mathcal{\hat E}_{k,0}$.
The definition of canonical momenta
$p_i \equiv {\partial \tilde L_2^{(0)}}/{\partial \dot{x_i}}$
does not determine velocities $\dot{x_i}$ as functions of $p_i$ and
$x_j$, but gives relations between $p_i$ and $x_j$:
%
%%%
\begin{equation}
\label{Eq:constraint}%5e
\tilde \varphi_i\equiv p_i + \frac{1}{2}\tilde G\epsilon_{ij}x_j =
0.
\end{equation}
%%%
%
According to Dirac's formalism of quantizing a constrained system,
such relations
%$\hat \varphi_i$
are primary constraints~\cite{Baxt,JZZ96}.
%~\cite{Baxt,JZZ96,HJWM}.
%~\cite{Baxt,JZZ96}.
%~\cite{JZZ96}.
%~\cite{JZZ96,HJWM}.
%
%Corresponding to
%The constraints are
%%%%
%%\begin{equation}
%%\label{Eq:constraint}%e
%$\hat \varphi_i\equiv p_i + \tilde G\epsilon_{ij}x_j/2 = 0$ (Eq.(19)
%of \cite{JZZ08}).
%%\end{equation}
%%%%
%
%Dynamics of this constrained system are solved by Eqs~(8)-(35) of
%Ref.~\cite{JZZ08}.
%
%The Lagrangian corresponding to the Hamiltonian $\hat H_2$ in
%Eq.~(\ref{Eq:CS-H1}) is
%%%%
%\begin{equation}
%\label{Eq:L2}%8e
%\hat L_2 = \frac{1}{2}M\dot{x_i}\dot{x_i} +
%%
%\frac{1}{2}G\epsilon_{ij}\dot{x_i}x_j- \frac{1}{2}K x_i x_i.
%\end{equation}
%%%%
%%
%%
%The reduced Lagrangian corresponding to $H_2^{(0)}$ is
%%%%
%\begin{equation}
%\label{Eq:L20}%9e
%\hat L_2^{(0)}= - \frac{1}{2}G\epsilon_{ij}\dot{x_i}x_j
%%
%- \frac{1}{2}K x_i x_i - \hat {\mathcal{E}}_{k,0}.
%\end{equation}
%%%%
%%\cite{JZZ08}.
%%%%%%%
%The momenta defined from $\hat L_2^{(0)}$ is
%%
%$p_i=\partial\hat L_0/\partial\dot{x_i}
%%
%= - G\epsilon_{ij}x_j/2$
%%
%which gives constraints
%%%%
%\begin{equation}
%\label{Eq:constraint}%10e
%\hat \varphi_i(x,p)=p_i + \frac{1}{2}G\epsilon_{ij}x_j.
%\end{equation}
%%%%
%% DDD which are the same as obtained from the reduced $L_0$
%%Eq.~(5) in the massless limit of~\cite{DJT}. DDD
%%%%%%%%%%%%%%%%%%%%%%%%%%%%%%%%%%%%%%
Because the Poisson brackets
$\{\tilde \varphi_i,\tilde \varphi_j\}_P= \tilde G\epsilon_{ij}\ne
0$,
the Dirac brackets are determined,
$\{x_i,p_j\}_D=\delta_{ij}/2$, ect.
The constraints $\tilde \varphi_i$ are strong conditions. They are
used to eliminate dependent variables:
%Thus the system has the reduced degrees of freedom:
four
%canonical
variables $(x_i,p_i), (i=1,2)$ are reduced to two independent ones
(e.g. $x_1,p_1$).
%%%%%%
Using these constraints to eliminate dependent variables,
the corresponding quantum commutators of independent variables
$\tilde x\equiv\sqrt{2}x_1$ and $\tilde p\equiv\sqrt{2}p_1$ are
$[\tilde x,\tilde p]=i\hbar$, ect.
%
%$\hat {\mathcal{E}}_{k,0}$ is a harmonic potential of an effectve
%constant $K$ (Eq.~(5) of Ref.~\cite{JZZ08})).
%
Then $\tilde H_2^{(0)}$ is rewritten as 1-dimensional harmonic
Hamiltonian plus $\hat {\mathcal{E}}_{k,0}$.
%Eq.~(25) of \cite{JZZ08})).
%
%$\hat H_0^{\ast}=\frac{1}{2\hat\mu^{\ast}}p^2 +
%%
%\frac{1}{2}\hat \mu^{\ast}\hat \omega^{\ast2}x^2+\mathcal{\hat
%E}_{k,0}$.
%with the effective mass and frequency.
%$\mu^{\ast}$ and frequency $\omega^{\ast}$:
%
%$\hat H_2^{(0)}=p^2/2\mu^{\ast} +
%%
%\mu^{\ast}\omega^{\ast2}x^2/2+\mathcal{E}_{k,0}$.
%%
The full Hamiltonian $\tilde H_2$ has two harmonic
modes~\cite{JZZ08,JZZ04b}.
%
%As is well-known,
The reduction to the reduced phase space
%the reduction $\hat H_2\to \hat H_2^{(0)}$
alters the symplectic structure. It leads to one mode of $\tilde
H_2$ going to infinity,
%essentially
decoupling from the system, and only one mode $\tilde H_2^{(0)}$
surviving.~\footnote
%\cite{H-limit}.
%%%%%%%%%%%%%%%%%%%%%%%%%%%%%%%%%%%%%%%%%%%%%%%%%%%%%%%%%%%%%%%%%%%%
{We compare dynamics in the present reduction and the reduction in
 the massless limit of \cite{DJT}.
 Lagrangian $\tilde L_2$,
 %$\hat L_2$ Eq.~(8),
 reduced $\tilde L_2^{(0)}$
 %Eq.~(17)
 and constraints $\tilde \varphi_i$
 %Eq.~(19) of \cite{JZZ08}
 are similar to Lagrangian
 $L$ Eq.~(1), reduced $L_0$ Eq.~(5) and constraints $C^i$ Eq.~(17) of
 \cite{DJT}.
 The reduction $\tilde L_2 \to \tilde L_2^{(0)}$ is similar to the
 reduction $L \to L_0$ of \cite{DJT}.
 %
 %In both $\hat H_{k,2}\to \hat {\mathcal{E}}_{k,0}$ and massless
 %limits,
 In both reductions, therefore the similar Chern- Simons type
 behavior and truncated states decoupling are obtained.}
% DDD In both massless and $H_{k,2}\to \mathcal{E}_{k,0}$ limits the
%constraints are the same, therefore the similar Chern-Simons type
%behavior and truncated states decoupling are obtained. DDD
%
%%%%%%%%%%%%%%%%%%%%%%%
%DDD
%
%
%~\cite{H-limit}.
%
% DDD In both massless and $H_{k,2}\to \mathcal{E}_{k,0}$ limits the
%constraints are the same, therefore the similar Chern-Simons type
%behavior and truncated states decoupling are obtained. DDD
%
%%%%%%%%%%%%%%%%%%%%%%%
%
%
% DDD
%In this limit the reduced system is a constrained one.
%%
%The point is that in this limit the definition of the canonical
%momenta from the corresponding reduced Lagrangian does not determine
%velocities $\dot{x_i}$ as functions of $x$ and $p$, but gives
%relations among $x$ and $p$.
%%
%According to Dirac's formalism of quantizing constrained system,
%such relations are primary constraints \cite{Baxt,JZZ96}.
%%
%Dynamics of this constrained system are solved by Eqs~(8)-(35) of
%Ref.~\cite{JZZ08}.
% DDD
%
%
% DDD These constraints
%$\varphi_i$
%are strong conditions. They can be used to eliminate dependent
%variables:
%%Thus the system has the reduced degrees of freedom:
%the four canonical variables $(x_i,p_i), (i=1,2)$ are reduced to two
%independent ones.
%%
%The Poisson brackets of the constraints are non-zero, the Dirac
%brackets of the canonical variables can be determined, then the
%corresponding quantum commutators are obtained.
%%
%As is well-known, the reduction to the reduced phase space alters
%the symplectic structure.
%%%%%%%%%%%%%
%As a consequence,
$\tilde H_2^{(0)}$ has a reduced set of eigenstates,
%it leads to a reduced set of eigenstates of the Hamiltonian $\hat
%H_2^{(0)}$,
and the eigenvalues
%$\mathcal{\tilde{J}}_{n}$ of
%the deformed angular momentum
of $\hat J_z$ then become
\begin{subequations}
\label{Eq:Jn}%6e
\begin{align}
\mathcal{\tilde{J}}_{n} & =\hbar \mathcal{\tilde{J}}
(n + \frac{1}{2}),\;\;(n = 0,1,2,\cdots)
\label{Eq:Jn.a}\\
\mathcal{\tilde{J}} & = 1 - \frac{m_{I}\omega_c\theta}{4\hbar}
- \frac{\eta}{m_I\omega_c\hbar + m_{I}^{2}\tilde
\omega_{\rho}^{2}\theta + \eta}.
%
%=\frac{1}{2},\;(\tilde{n}=0,1,2,\cdots)
%
\label{Eq:Jn.b}
\end{align}
\end{subequations}
%%%
Where the two terms $O(\theta)$ and $O(\eta)$ are corrections due to
the spatial noncommutativity, which are inaccessible to experiment
because they are vanishingly small.

In the
%present
case of both position-position and momentum-momentum noncommuting,
there is an effective intrinsic magnetic field
$B_{eff}\sim\eta$~\cite{JZZ08}.
Thus a further limiting process of diminishing the external magnetic
field $B$ ($\omega_c$) to zero is meaningful, and the surviving
system has non-trivial dynamics.
In this limit we have
$\eta/(m_I\omega_c\hbar+m_I^2\tilde \omega_{\rho}^2\theta + \eta)\to
\eta/(m_I^2\tilde \omega_{\rho}^2\theta + \eta)$.
Using the consistency condition~\footnote
%
%\cite{proportion},
%%%%%%%%%%%%%%%%%%%%%%%%%%%%%%%%%%%%%%%%%%%%%%%%%%%%%%%%%%%%%%%%%%%%
{The proportionality of the NC parameters
 $\theta$ and $\eta$ is determined by fundamental principles.
 %~\cite{BRACZ,BRACZb}.
 %has been questioned in Refs.~\cite{BRACZ,BRACZb}.
 At the quantum mechanics level, the general structures of the deformed
 annihilation and creation operators which satisfy a complete and
 closed deformed bosonic algebra at the non-perturbation level were
 obtained in Ref.~\cite{JZZ08b}. The proportionality $\eta=K\theta$
 between the NC parameters $\theta$ and $\eta$ is clarified from the
 consistency of the deformed Heisenberg-Weyl algebra with the
 deformed bosonic algebra. $\theta$ is a fundamental constant.
 $K$ depends on some dynamical parameters of Lagrangian.
 %of the considered system.
 %
 From the definition of momenta being the partial derivatives of
 Lagrangian with respect to the NC coordinates, the dependence of
 $\eta$ on the dynamical parameters of the considered system is
 understood.}
%%%%%%%%%%%%%%%%%%
%
$\eta = m_I^2\tilde \omega_{\rho}^2\theta$,
this leads to a cancelation
%near-cancelation
between the NC parameters $\theta$ and $\eta$ so that this term
equals $1/2$, and
$\mathcal{\tilde{J}} = 1/2 - m_{I}\omega_c\theta/4\hbar$, where
$1/2$ dominates $\mathcal{\tilde{J}}$.
Therefore, the dominant value of the zero-point angular momentum
$\mathcal{\tilde J}_0$ assumes a fractional value:~\footnote
%%%%%%%%%%%%%%%%%%%%%%%%%%%%%%%%%%%%%%%%%%%%%%%%%%%%%%%%%%%%%%%%%%%%%%%%%%
{There is a subtle point related to taking the meaningful limits
 $\theta,\eta\rightarrow 0$ and $B\rightarrow 0$.
 %
 %are subtle.
 %
 In the limits
 $\theta,\eta\rightarrow 0$, deformed dynamics in NC space
 %(Eqs~(8)-(35) of~\cite{JZZ08})
 is reduced to undeformed one in commutative space.
 The reduced system $\tilde H_2^{(0)}$ is a constrained one.
 The Deformed Poisson brackets of the constraints are
 $\{\tilde \varphi_i,\tilde \varphi_j\}_P= \tilde G\epsilon_{ij}$.
 %(Eq.~(22) of~\cite{JZZ08}) of constraints
 %Eq.~(\ref{Eq:constraint})
 %(Eq.~(22) of~\cite{JZZ08})
 %%
 %$\{\hat \varphi_i,\hat \varphi_j\}_P= G\epsilon_{ij}$
 %
 In the limits $\theta,\eta\rightarrow 0$, they are reduced to
 undeformed ones in commutative space,
 $\{\varphi_i,\varphi_j\}_P= m_I\omega_c\epsilon_{ij}$.
 If we followed with
 %the limit
 $B\rightarrow 0 (\omega_c\rightarrow 0)$, we would obtain
 $\{\varphi_i,\varphi_j\}_P=0$, thus Dirac brackets of canonical
 variables would not be determined, and the system would not survive
 at the quantum level.
 This indicates that in Eq.~(\ref{Eq:Jn.b}) when we take
 %the limits
 $\theta,\eta\rightarrow 0$ first to yield the conventional result,
 it makes no sense to follow with
 %the limit
 $B\rightarrow 0$.
 On the other hand, if we take
 %the limit
 $B\rightarrow 0$ first, the deformed Poisson brackets are reduced to
 $\{\varphi_i,\varphi_j\}_P=
 (m_I^2\omega_{\rho}^2\theta+\eta)\epsilon_{ij}/\hbar$.
 This shows that the subsequent limit $\theta,\eta\rightarrow 0$ also
 is meaningless.
 Therefore, {\it only in NC space} non-trivial dynamics of the
 reducced system $\tilde H_2^{(0)}$ survives at the quantum level in
 the limit $B\rightarrow 0$.}
%%%%%%%
$\hbar/4$.
This is a distinct NC signal, which is within the limits of
measurable accuracy of current experiments.

%\vspace{0.5cm}

%%%%%%%%%%%%%%%%%%%%%%%%%%%%%%%%%%%%%%%%%%%%%%%%%%%%%%%%%%%%%%%%%%%%%   aaa
{\bf 4. MHFS induced by FAM $\mathcal{\tilde{J}}_{0}$} --
%
%\vspace{0.5cm}
%
We consider a doubly-charged alkaline-earth ion $I^{++}$ caught in a
combined-field trap. The trapping mechanism is provided by a uniform
magnetic field ${\bf B}$ aligned along the $z$-axis and an
electrostatic potential~(\ref{Eq:V}).
For an alkaline-earth atom, the outer subshell has two $s$
electrons, and the inner shells are completely filled. When the two
$s$ electrons of the outer shell are ionized, the resulting
double-ion $I^{++}$ also has rotational symmetry and resembles an
effective spherical nucleus. We consider an electron injected into
the trap and the captured electron together with this ion forms a
singly-charged ion $I^{+}$ which is still stably trapped.
It is required that the principal quantum number $n$ of the captured
electron is large enough so that the system is a Rydberg one.
In a reasonable approximation, the energy spectrum of the Rydberg
electron is calculated on a similar basis as for a hydrogen-like
system.

According to the above analysis, in the case where both
position-position and momentum-momentum operators are noncommuting,
and under the aforementioned conditions, the trapped ion $I^{++}$
possesses FAM $\mathcal{\tilde J}_0$. Correspondingly, there is a
zero-point magnetic momentum ${\tilde \mu}_0$,
\begin{equation}
\label{Eq:mu}%7e
%\tilde \mu}_0 = Z^{*}e\mathcal{\tilde J}_0/2m_I
%%
%= Z^{*}\mu_N\mathcal{\tilde J}_0/A\hbar,
%
{\tilde \mu}_0 = \frac{Z^{*}e}{2m_I}\mathcal{\tilde J}_0
= \frac{Z^{*}\mu_N}{A\hbar}\mathcal{\tilde J}_0,
\end{equation}
where $m_I=Am_P$ ($m_P$ is proton mass and $A$ is nuclear mass
number), $\mu_N=e\hbar/2m_p$ is nuclear magneton.

The magnetic interaction between the magnetic momentum ${\tilde
\mu}_0$ and the magnetic fields of the Rydberg electron induces
magnetic hyperfine structures of the energy spectrum of the Rydberg
electron.
Thus the measurement of FAM $\mathcal{\tilde J}_0$, through the
corresponding
%zero-point magnetic momentum
${\tilde \mu}_0$, is turned into measuring MHFS of the Rydberg
electron.
Similar to
%the magnetic hyperfine structures
MHFS generated by nuclear spin~\cite{Slat,Yang}, splitting of MHFS
induced by $\mathcal{\tilde J}_0$ of the ion $I^{++}$ can be
calculated in two equivalent approaches~\cite{Slat}: investigating
the interaction of the ion $I^{++}$ on the Rydberg electron, or
discussing the equivalent interaction of the Rydberg electron on the
ion $I^{++}$. In the following, we apply the second approach.

To get a clean signal of such induced MHFS, we choose some even-even
nucleus, because the nuclear spin of an even-even nucleus is zero.

%\vspace{0.5cm}

%%%%%%%%%%%%%%%%%%%%%%%%%%%%%%%%%%%%%%%%%%%%%%%%%%%%%%%%%%%%%%%%%%%%%
\textit{The magnetic hyperfine interaction}~\cite{Slat,Yang} --
In the center of mass system the magnetic hyperfine splitting of the
energy spectrum of the Rydberg electron induced by $\mathcal{\tilde
J}_0$ of the ion $I^{++}$ is described by
the effective hyperfine interaction Hamiltonian $H^{(hfs)}_{in}$
between
$\vec{\tilde \mu}_0
%
%=-(Z^{*}\mu_N/A\hbar){\mathcal{\vec{\tilde J}}}_0
%
=-(Z^{*}\mu_N/A\hbar)(0,0,{\mathcal{\tilde J}}_0)$
of the ion $I^{++}$ and the magnetic fields generated at the
position of the ion $I^{++}$ by the Rydberg electron.
%%%
%\begin{align}
%H_{in}^{(hfs)}  &  =\frac{1}{4\pi \epsilon_{0}c^{2}}
%%
%\frac{Z^{\ast}\mu_{N}}{A\hbar}\frac{2\mu_{B}}{\hbar}
%%
%\left[  \frac{8\pi}{3}\delta(\mathbf{r})
%%
%+ \frac{l(l+1)}{j(j+1)}\langle \frac{1}{r^{3}}\rangle_{av}\right]
%%
%\nonumber \\
%&  \times{\mathcal{\vec{\tilde{J}}}}_{0}\cdot{\vec{j}},
%\label{Eq:H-hfs}%13e
%\end{align}
%%%%
%where $\mu_B$
%%$\mu_B = e\hbar/2m_e$
%is Bohr magneton, $\vec j$
%%$\vec j = \vec l + \vec s$
%is the total angular momentum of the electron and $\langle 1/r^3
%\rangle_{av}$ is the average value of $1/r^3$ in the corresponding
%state.
%
The corresponding splitting and intervals of the electronic energy
spectrum are
$\Delta E_{nljm_{j}}^{(hfs)}
= \langle nljm_{j}|H_{in}^{(hfs)}|nljm_{j} \rangle
=A_{nlj}{\mathcal{\tilde{J}}}_{0}m_{j}\hbar$, $\;$
$\Delta E_{nlj}^{(hfs)}(\Delta m_{j})
\equiv \Delta E_{nljm_{j}}^{(hfs)}-\Delta E_{nljm_{j^{\;
\prime}}}^{(hfs)}
=A_{nlj}{\mathcal{\tilde{J}}}_{0}\Delta m_{j}\hbar$,
%
%%%
%\begin{subequations}%e
%\begin{align}
%\Delta E_{nljm_{j}}^{(hfs)} &  =\langle nljm_{j}|H_{in}^{(hfs)}|nljm_{j}%
%\rangle \nonumber \\
%&  =A_{nlj}{\mathcal{\tilde{J}}}_{0}m_{j}\hbar,
%%
%\label{Eq:dE-nlj.a}\\
%%
%\Delta E_{nlj}^{(hfs)}(\Delta m_{j}) &  =\Delta
%E_{nljm_{j}}^{(hfs)}-\Delta
%E_{nljm_{j^{\; \prime}}}^{(hfs)}\nonumber \\
%&  =A_{nlj}{\mathcal{\tilde{J}}}_{0}\Delta m_{j}\hbar,
%
%\label{Eq:dE-nlj.b}                                           %
%%
%\end{align}
%
%\end{subequations}
%%%
where $\Delta m_j=m_j-m_{j^{\;\prime}}$.

We consider the even-even nucleus of Magnesium ($Z=12, A=24$). When
two $s$ electrons at the $M$ shell are ionized, the ion $Mg^{++}$
has a spherical configuration. The Rydberg electron should fills
shells of $n > 3$.
We estimate the magnetic hyperfine splitting
%MHFS
and intervals of the spectrum of the Rydberg electron of $n=6, l=0$.

%\vspace{0.5cm}

%%%%%%%%%%%%%%%%%%
%\begin{quote}
%I) $l=0$\emph{ case}
%\end{quote}
%
For an $s$ electron, $l=0, j=1/2, m_j=\pm 1/2,\Delta m_j=1$.
Owing to the non-vanishing electronic charge density at the ion
$Mg^{++}$, the only contribution to the Hamiltonian $H^{(hfs)}_{in}$
comes from the Fermi contact interaction.
%
%The corresponding
%$A_{nlj}$ reads
From
$A_{n0\frac{1}{2}}
= (8/3)(m_{e}/Am_{p})\alpha^{4}(m_{e}c^{2})
(Z^{\ast}/n)^{3}/\hbar^{2}$,
%
%$A_{n0\frac{1}{2}}
%%
%= (8/3)(m_{e}/Am_{p})\alpha^{4}(m_{e}c^{2})
%%
%(Z^{\ast}/n)^{3}/\hbar^{2}\\
%\sim 4.449\times 10^{-17}eV/\hbar^{2}$,
%
%%%%
%\begin{align}
%\label{Eq:A-s}%15e
%%
%A_{n0\frac{1}{2}}& =\frac{8\pi}{3}\frac{1}{4\pi \epsilon_{0}c^{2}}
%%
%\frac{Z^{\ast}\mu_{N}}{A\hbar}\frac{2\mu_{B}}{\hbar}
%%
%|\psi(0)|_{\;n0\frac{1}{2}}^{\;2}
%%
%\nonumber \\
%& =\frac{8}{3}\frac{m_{e}}{Am_{p}}\alpha^{4}(m_{e}c^{2})
%%
%\left(\frac{Z^{\ast}}{n}\right)  ^{3}\frac{1}{\hbar^{2}}.
%%
%\end{align}
%%%
%In the above $(A=24, \; Z^*=2, \; n=6)$,
%%
%%$|\psi(0)|_{\;n0\frac{1}{2}}$
%$|\psi(0)|^{\;2}_{\;n0\frac{1}{2}}={Z^*}^3/\pi a^3_1 n^3$ is the
%wavefunction of the electron at the nucleus,
%%
%%$|\psi(0)|^{\;2}_{\;n0\frac{1}{2}}={Z^*}^3/\pi a^3_1 n^3$,
%%
%$a_1$  is Bohr radius,
%%$a_1=4\pi\epsilon_0\hbar^2/m_e e^2$ (Bohr radius),
%%
%and $\alpha$  is fine-structure constant.
%%$\alpha=e^2/4\pi\epsilon_0\hbar c$ (fine-structure constant).
%
%
it follows that the magnetic hyperfine splitting and intervals have
orders
%
%%%%%%%%%%%%%%%%%%%%%%%%%%%%%%%%%%%%%%%%%%%%%%%%%%%%%%%%%%%%%%%
%$\Delta E_{60\frac{1}{2}\frac{\pm1}{2}}^{(hfs)}
%%
%=\pm\frac{1}{2}A_{60\frac {1}{2}}{\mathcal{\tilde{J}}}_{0}\hbar
%%
%\sim \pm5.5\times10^{-8}\text{eV}$, $\:$
%%
%$\Delta E_{60\frac{1}{2}}^{(hfs)}(1)
%%
%=\Delta E_{60\frac{1}{2}\frac{1}{2}}^{(hfs)}
%%
%-\Delta E_{60\frac{1}{2}\frac{-1}{2}}^{(hfs)}
%%
%\sim1.1\times10^{-7}\text{eV}$.
%%%%%%%%%%%%%%%%%%%%%%%%%%%%%%%%%%%%%%%%%%%
\begin{subequations}%8e
\begin{align}
\Delta E_{60\frac{1}{2}\frac{\pm1}{2}}^{(hfs)}
& = \pm\frac{1}{2}A_{60\frac {1}{2}}{\mathcal{\tilde{J}}}_{0}\hbar
\sim \pm 5.5\times10^{-8} eV,
\label{Eq:dE-601.a}\\
\Delta E_{60\frac{1}{2}}^{(hfs)}(1)
&  =\Delta E_{60\frac{1}{2}\frac{1}{2}}^{(hfs)}
-\Delta E_{60\frac{1}{2}\frac{-1}{2}}^{(hfs)}
%
%\nonumber \\
\sim1.1\times10^{-7} eV.
\label{Eq:dE-601.b}
\end{align}
\end{subequations}

Measurements of
$\Delta E^{(hfs)}_{nljm_j}$ and/or
$\Delta E^{(hfs)}_{nlj}(\Delta m_j)$ directly determine FAM
$\mathcal{\tilde J}_0$,
%
%$\mathcal{\tilde J}_0
%%
%(= \Delta E_{nljm_{j}}^{(hfs)}/A_{nlj}m_{j}\hbar
%%
%= \Delta E_{nlj}^{(hfs)}(\Delta m_{j})/A_{nlj}\Delta m_{j}\hbar)$,
%
thus providing signals of spatial noncommutativity.

{\bf 5. Testing  spatial noncommutativity via MHFS by FAM
$\mathcal{\tilde J}_0$} --
%
%\vspace{0.5cm}
%
An ionic core with a closed shell configuration such as $Mg^{++}$ is
a conceptually ideal system to testing spatial noncommutativity.
$Mg^{++}$ is trapped by a combination of an electrostatic
potential~(\ref{Eq:V}) and a uniform magnetic field ${\bf B}$
aligned along the $z$-axis~\cite{JZZ08}.
%
%The limit of $\hat H_{k,2}$
%%the mechanical kinetic energy
%of
%%the ion
%$Mg^{++}$ approaching $\hat {\mathcal{E}}_{k,0}$
%lowest eigenvalue
According to the mentioned approach, the reduced system $\hat
H_{2}^{(0)}$ is realized.
%~\cite{Sten,WI97}.
%~\cite{H-limit}.
%
In the well defined limits, the surviving system has non-trivial
dynamics, and FAM of
%the ion
$Mg^{++}$ is $\mathcal{\tilde J}_0$.
%$\mathcal{\tilde J}_0 = \hbar/4$.
%
To make $\mathcal{\tilde J}_0$
%this angular momentum
observable,
%in the experiment,
we inject an electron into the trap, capturing it in a high Rydberg
state of an appropriate principal quantum number $n$ by
%the ion
$Mg^{++}$.
The coupling between
%the zero-point magnetic momentum
${\tilde \mu}_0$ of
%the ion
$Mg^{++}$ and the magnetic fields generated at the position of
%the ion
$Mg^{++}$ by the Rydberg electron will induce the magnetic hyperfine
splitting of the electronic energy spectrum which
%%%%%%
%In this way, t
%These magnetic hyperfine splitting and intervals of the electronic
%spectrum
is signal of spatial noncommutativity. Their orders are
%of these splitting are
%
$\sim 10^{-7} - 10^{-8} eV$,
%$10^{-8}\text{eV}$$\sim$$10^{-9}\text{eV}$,
which lie within the limits of measuring accuracy of current
experiments. This experiment can be achieved by existing technology,
for example, high-resolution laser spectroscopy. Considering the
pollution from other interactions during the measurement, we should
pick up the true signal contributing the magnetic hyperfine
splitting induced by FAM. This is achieved by the experiments which
are performed twice: one with the magnetic field detuned to zero and
one without the detuning process.

%\vspace{0.5cm}

%%%%%%%%%%%%%%%%%%%%%%%%%%%%%%%%%%%%%%%%%%%%%%%%%%%%%%%%%%%%%%%%%%%%%%
{\bf 6. Summary} --
%
%\vspace{0.5cm}
%
NCQM is a candidate of possible new physics.
%one of the current candidates in tracking down new physics.
%
At first sight, it seems that NCQM is unverifiable.
However, we found that MHFS induced by FAM is one of the most
important effect of spatial noncommutativity which, under
well-defined conditions, lies within the range of normal laboratory
measurements.
Physics beyond the standard model is speculative. Its experimental
tests are the focus of broad interest, especially the MHFS approach
is reasonable achievable with current technology.
Comparing with the experiments performed on quasi-bound Rydberg
states in crossed fields~\cite{CHLZ}, via a Chern-Simons
process~\cite{JZZ06}, using modified electron momentum
spectroscopy~\cite{JZZ08} and others, MHFS is the most effective
approach.
Based on the unique feature of the MHFS approach, its experimental
observation will be a key step towards a decisive test of confirming
or ruling out spatial noncommutativity.

%\vspace{0.5cm}

%%%%%%%%%%%%%%%%%%%%%%%%%%%%%%%%%%%%%%%%%%%%%%%%%%%%%%%%%%%%%%%%%

\vspace{0.5cm}

JZZ's work is supported by NSFC (No. 10575037) and SEDF; LKG's work
is supported by NSFC (No. 60490280) and NFRPC (No. 2005CB724500);
HPL's work is supported by NSFC (No. 10774162); WC's work is
supported by the Swiss National Foundation (Grant No.
200020-125124).

%%%%%%%%%%%%%%%%%%%%%%%%%%%%%%%%%%%%%%%%%%%%%%%%%%%%%%%%%%%%%%%%%

\end{document}